\documentclass[aps,prb,reprint,superscriptaddress]{revtex4-1}
\usepackage{amsmath}
\usepackage{amssymb}
\usepackage{graphicx}
\graphicspath{{./figures/}}

\newcommand{\mvec}[1]{\ensuremath{\mathbf{#1}}}

\newcommand{\Heff}{\ensuremath{\mvec{H}_\mathrm{eff}}}

\begin{document}
\title{Microwave-induced dynamic switching of magnetic skyrmion cores in nanodots}

\begin{abstract}
The nonlinear dynamic behavior of a magnetic skyrmion in circular nanodots was studied numerically
by solving the Landau-Lifshitz-Gilbert equation with a classical spin model.
We show that a skyrmion core reversal can be achieved within nanoseconds using a perpendicular oscillating magnetic field.
Two symmetric switching processes that correspond to excitations of the breathing mode and the mixed mode (combination of the breathing mode and a radial spin-wave mode)
are identified.
For excitation of the breathing mode,
the skyrmion core switches through nucleation of a new core from a transient uniform state.
In the mixed mode, the skyrmion core reverses with the help of spins excited
both at the edge and core regions. Unlike the magnetic vortex core reversal, the excitation of radial spin waves
does not dominate the skyrmion core reversal process.
\end{abstract}


\author{Bin Zhang}
\affiliation{Institut f\"{u}r Experimentalphysik, Freie Universit\"{a}t Berlin, Arnimallee 14, 14195 Berlin, Germany}
\author{Weiwei Wang}
\affiliation{Faculty of Engineering and the Environment, University of Southampton,  SO17 1BJ, Southampton, United Kingdom}
\author{Marijan Beg}
\affiliation{Faculty of Engineering and the Environment, University of Southampton,  SO17 1BJ, Southampton, United Kingdom}
\author{Hans Fangohr}
\affiliation{Faculty of Engineering and the Environment, University of Southampton,  SO17 1BJ, Southampton, United Kingdom}
\author{Wolfgang Kuch}
\email{kuch@physik.fu-berlin.de}
\affiliation{Institut f\"{u}r Experimentalphysik, Freie Universit\"{a}t Berlin, Arnimallee 14, 14195 Berlin, Germany}
\maketitle


A rich variety of magnetic configurations can emerge in nanoscale magnetic materials. For instance, in a circular nanodisk, both vortex and skyrmion states
may occur.~\cite{Cowburn1999, VanWaeyenberge2006, Kim2014, Im2012}
As a topological excitation, a skyrmion has a topological charge $Q=\pm 1$, whereas the vortex topological charge \cite{Braun2012}
is $\pm 1/2$. The reversal of a vortex core has been intensively studied in the past few years~\cite{Schneider2001, 
Guslienko2001, Okuno2002, Thiaville2003, VanWaeyenberge2006, Xiao2006, Hertel2007, Guslienko2008, Yoo2012, Im2012, Pylypovskyi2013, Ruckriem2014}
 because of its potential use in data storage devices and the importance in fundamental physics.
Recently, the creation and annihilation of individual skyrmions was demonstrated,~\cite{Tchoe2012, Finazzi2013, Iwasaki2013, Romming2013,Lin2013} which illustrates
the potential application of topological charge in future information-storage devices.
Moreover, the spin-wave modes of a skyrmion have been investigated both experimentally~\cite{Okamura2013, Onose2012} and using simulations,~\cite{Mochizuki2012, Kim2014}
and three $k=0$ skyrmion ``optical'' modes have been identified. It was found that anticlockwise
and clockwise rotation modes are excited when an alternating current (AC) magnetic field is applied in the skyrmion plane,
while the breathing mode is found if the AC field is applied perpendicular to the skyrmion plane.~\cite{Mochizuki2012}
Very recently, a hysteretic behavior and reversal of an isolated skyrmion formed in a nanodisk by a static magnetic field have been demonstrated.~\cite{Beg2014}  However, the influence of a time-dependent magnetic field, such as a microwave field, on the skyrmion core reversal is still unexplored.  This is the focus of this letter, in which we study the reversal of an individual skyrmion in a circular magnetic dot by applying an AC magnetic field perpendicular to the skyrmion plane.

The Dzyaloshinskii-Moriya
interaction (DMI)~\cite{Dzyaloshinskii1958, Moriya1960} is an antisymmetric interaction
which arises if inversion symmetry in magnetic systems is absent,~\cite{Fert2013,Nagaosa2013} either because of a non-centrosymmetric crystal
structure or due to the presence of interfaces.~\cite{Moon2013} 
Consequently, these chiral interactions can be classified
as bulk or interfacial, depending on the type of inversion symmetry breaking.~\cite{Moon2013}
In magnetic thin films, a Bloch-type chiral skyrmion can be formed by bulk DMI, while a N\'{e}el-type radial (``hedgehog") skyrmion requires the presence of interfacial DMI.~\cite{Zhou2014} In this study we consider the bulk DMI, which exists in non-centrosymmetric magnets such as MnSi~\cite{Muhlbauer2009, Schulz2012} and FeGe.~\cite{Huang2012}

\begin{figure}[tbhp]
\begin{center}
\includegraphics[scale=0.45]{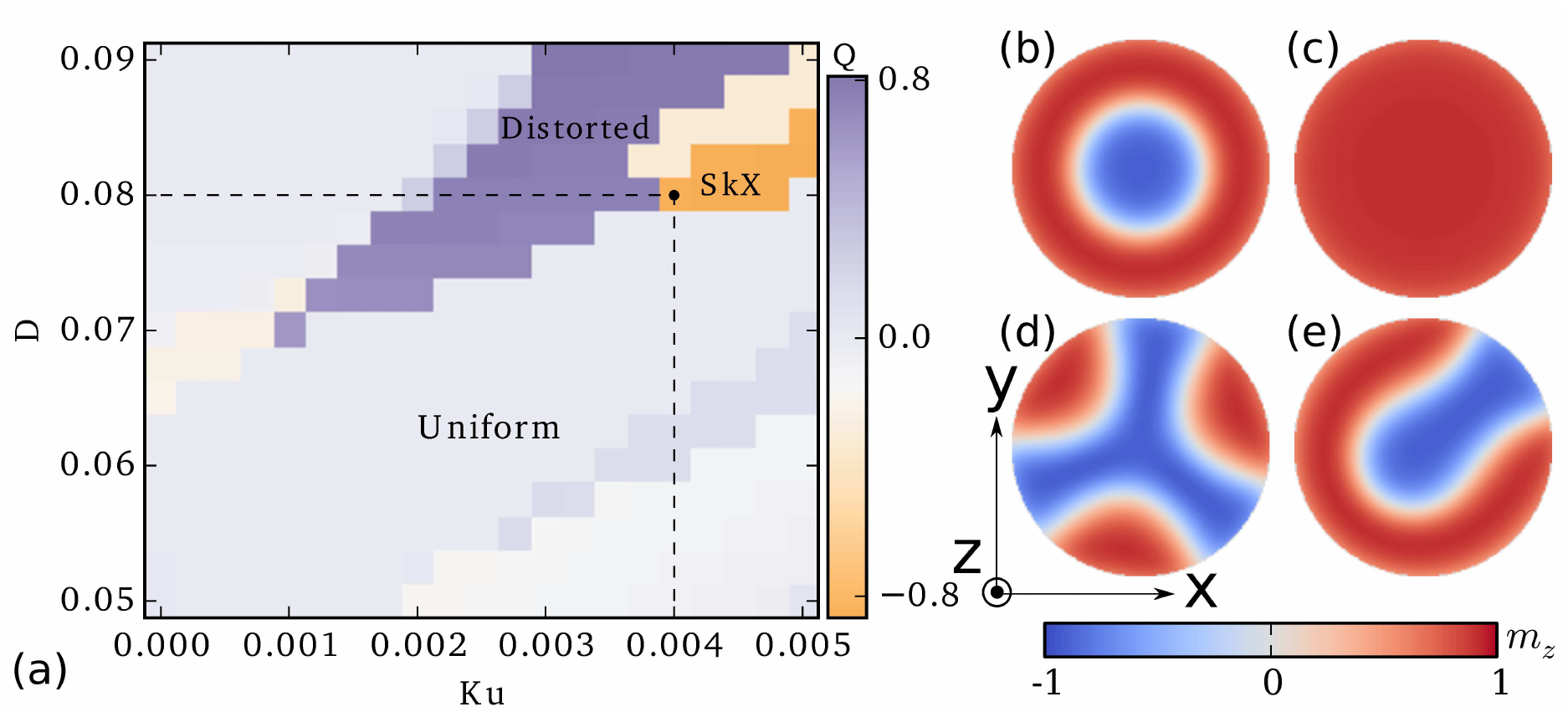}
\caption{(a) The phase diagram with topological charge of the ground state as a function of $D$ and $K_u$.
Possible ground states are (b) skyrmion (SkX), (c) uniform, and (d)--(e) distorted states. The diameter of the dot is 121 sites.}
\label{fig1}
\end{center}
\end{figure}

We consider a classical Heisenberg model on a two-dimensional regular square lattice with ferromagnetic exchange interaction (represented by $J$) and uniaxial anisotropy ($K_u$) along the $z$ axis.~\cite{Yi2009, Mochizuki2012, Kong2013}  Apart from that, a time-dependent magnetic
field $\mvec{h}(t)$ is applied to the system in the positive $z$ direction.
Therefore, the total Hamiltonian of the system is given by
\begin{align}\label{eq_ham}
  \mathcal{H}=&-J \sum_{<i,j>} \mvec{m}_i\cdot \mvec{m}_{j} +\sum_{<i,j>} \mvec{D}_{ij}\cdot [\mvec{m}_i \times \mvec{m}_j]  \nonumber\\
  &- \sum_{i} K_u  (\mvec{e}_z \cdot \mvec{m}_i)^2   - \sum_{i} |\boldsymbol \mu_i|  \mvec{h} \cdot  \mvec{m}_i,
\end{align}
where $\mvec{m}_i$ is the unit vector of a magnetic moment $\boldsymbol \mu_i =-\hbar \gamma \mvec{S}$ with
$\mvec{S}$ being the atomic spin and $\gamma(>0)$ the gyromagnetic ratio.
The DMI vector $\mvec{D}_{ij}$ can be written as $\mvec{D}_{ij} = D \mvec{\hat{r}}_{ij}$ in the case of bulk DMI,
where $D$ is the DMI constant and $\mvec{\hat{r}}_{ij}$ is the unit vector between $\boldsymbol \mu_i$ and $\boldsymbol \mu_j$.
In this study, dipolar interactions are not included, because in systems with small sizes of 10-100 nm, dipolar interactions are comparably weak.~\cite{Rossler2006, Mochizuki2012}
The spin dynamics at lattice site $i$ is governed by the Landau-Lifshitz-Gilbert (LLG) equation,
\begin{align}\label{eq_llg}
  \frac{\partial \mvec{m}_i}{\partial t} = - \gamma \mvec{m}_i \times \Heff + \alpha \mvec{m}_i \times  \frac{\partial \mvec{m}_i}{\partial t},
\end{align}
where $\alpha$ denotes the Gilbert damping and the effective field $\Heff$ is computed as $\Heff = - \frac{1}{|\boldsymbol \mu_i|} \frac{\partial \mathcal{H}}{\partial \mvec{m}_i}$. The Hamiltonian~(\ref{eq_ham}) associated with the LLG equation~(\ref{eq_llg}) can be understood as a
finite-difference micromagnetic model. In this study, we have chosen $J=1$ as the energy unit.~\cite{Mochizuki2012}
All simulations are performed in a circular dot with a diameter of $121$ lattice sites.
We have chosen $J=\hbar=\gamma=S=1$ as simulation parameters, and therefore, the coefficients for conversion of the
external field $h$, time $t$, and frequency $\omega$ to SI units are $J/(\hbar \gamma S)$,
$\hbar S/J$ and $J/(\hbar S)$, respectively, as shown in Table~\ref{tab1}.
The Gilbert damping $\alpha=0.02$ is selected for most of the simulations.
A DMI constant with a value $D=0.08$ is used, except for Fig.~\ref{fig1}, where $D$ is varied.  This value of $D$ results in the spiral period $\lambda \approx 2 \pi J a/D \approx 39$ nm
and the skyrmion diameter~\cite{Mochizuki2012} of
approximately
55 nm for the lattice parameter $a=0.5$ nm.

\begin{table}
\caption{\label{tab1} Unit conversion table for $J=1$ meV and $S=1$.}
\begin{ruledtabular}
\begin{tabular}{lll}
Magnetic field $h$ & $J/(\hbar \gamma S)$ & $\approx 8.63$ T \\
Time $t$ & $\hbar S/J$ & $\approx 0.66$  ps \\
Frequency $\omega$ & $J/(\hbar S)$  & $\approx 1.52 \times 10^{3}$ GHz \\
\end{tabular}
\end{ruledtabular}
\end{table}

There are various stable states for the system described by Hamiltonian~(\ref{eq_ham}). The phase diagram presented in Fig.~\ref{fig1}(a) shows the ground state as a function of $D$ and $K_{u}$ for the magnetic nanodot. The phase diagram is obtained by comparing several possible states such as skyrmion, uniform, and distorted states, as shown in Figs.~\ref{fig1}(b)--(e). Figure~\ref{fig1}(b) shows an isolated skyrmion with topological charge (skyrmion number) $Q = \int q \mathrm{d}x \mathrm{d}y \approx-0.86$ where $q = \mvec{m} \cdot (\partial_x \mvec{m} \times \partial_y \mvec{m})$ is the topological charge density. The isolated skyrmion state emerges in the brown region with skyrmion number $\lesssim -0.6$, while the distorted states are shown in the purple area, where the skyrmion number is chosen to be positive to
distinguish from the skyrmion state. The uniform state with skyrmion number $\approx 0$ is found in the remaining area.
We fix parameters $D = 0.08$ and $K_u = 0.004$ in the remaining of this work, as indicated in Fig.~\ref{fig1}(a).

\begin{figure}[tbhp]
\begin{center}
\includegraphics[scale=0.35]{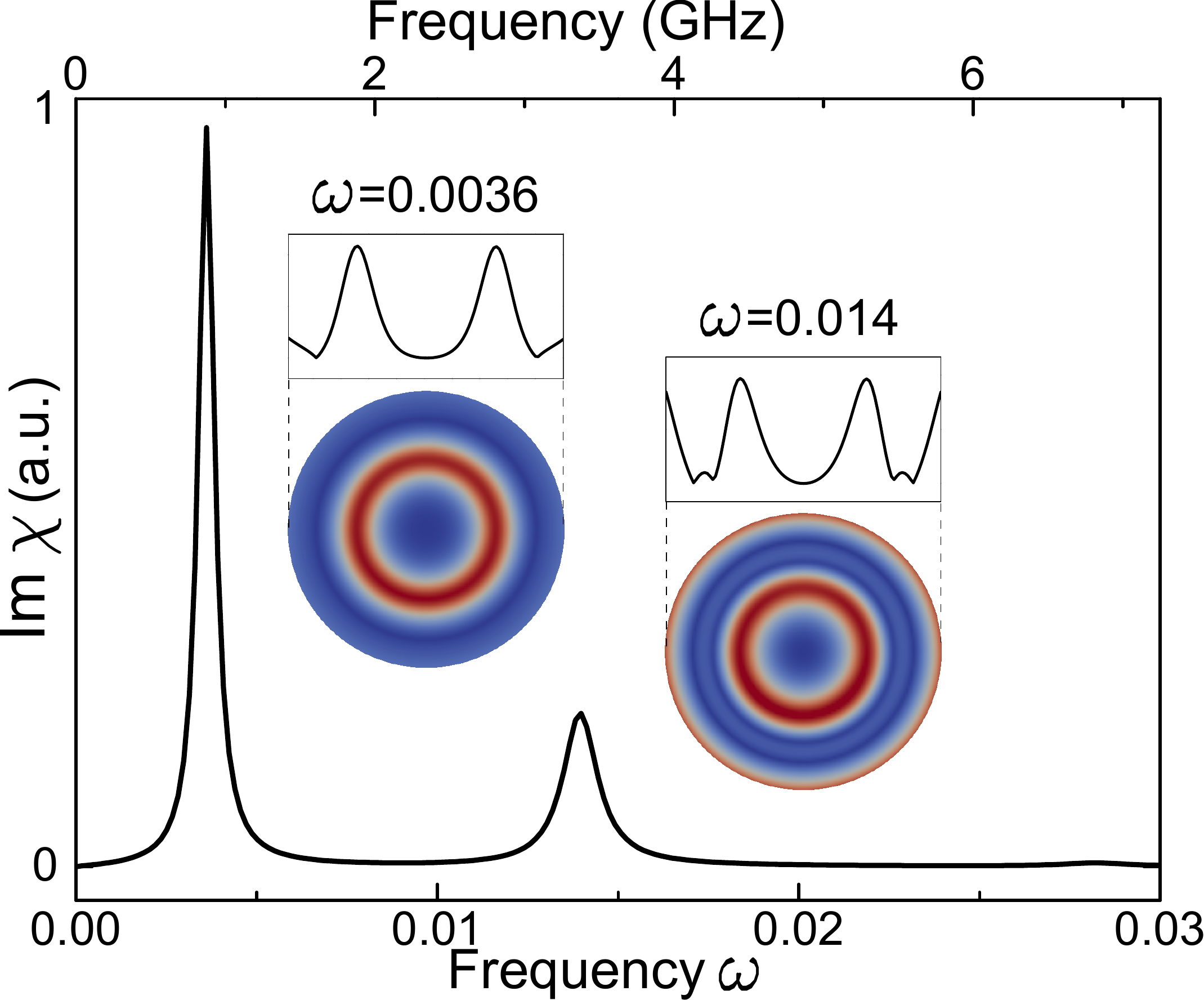}
\caption{Imaginary part of the susceptibility spectrum ($\chi_{zz}$) obtained after applying a field pulse $h_z = h_0 \mathrm{sinc}(\omega_0 t)$
along the $z$ axis to the system with $h_0=8\times10^{-6}$ and $\omega_0=0.1$. The insets show the spatial distribution of the FFT power
at eigenfrequencies $\omega=3.6 \times 10^{-3}$ and $1.4 \times 10^{-2}$, respectively.
The lines are profiles of the $m_z$ component across the core center.
}
\label{fig2}
\end{center}
\end{figure}

To extract the dynamic response of the skyrmion to the microwaves, we calculate the magnetic absorption spectrum of the skyrmion~\cite{Liu2008, Kim2014}
by applying a magnetic field pulse in the vertical ($z$) direction with a temporal shape given by a sinc-function field $h_z = h_0\; \mathrm{sinc}(\omega_0 t) = h_0\; \mathrm{sin}(\omega_0 t) / (\omega_0 t)$,
where $h_0=8\times10^{-6}$ and $\omega_0=0.1$.
The magnetic spectrum of a single skyrmion is shown in Fig.~\ref{fig2} with $\alpha=0.04$.
There are two main resonance lines at $\omega = 3.6 \times 10^{-3}$ and $\omega = 1.4 \times 10^{-2}$, corresponding to $f=0.87$ GHz and $f=3.39$ GHz, respectively,
if $J=1$ meV and $S=1$ are chosen.
The insets in Fig.~\ref{fig2} show the corresponding eigenmode amplitude plots, which are obtained by computing the fast Fourier transform (FFT)
to the spatial $m_z$ oscillations of the sample. These patterns indicate the spatially resolved fluctuation amplitude.
The lower-frequency resonance peak corresponds to the breathing mode of the skyrmion,~\cite{Mochizuki2012, Onose2012}
where the skyrmion radially expands and shrinks as a function of time with significant fluctuation
around the skyrmion core region.
The higher-frequency resonance peak with weaker intensity can be viewed as a mixed mode that combines
the breathing mode and the radial spin-wave mode,~\cite{Kim2014} since the spin excitation arises both around the core area as well as at the dot edges.

\begin{figure*}[tbhp]
\begin{center}
\includegraphics[scale=0.28]{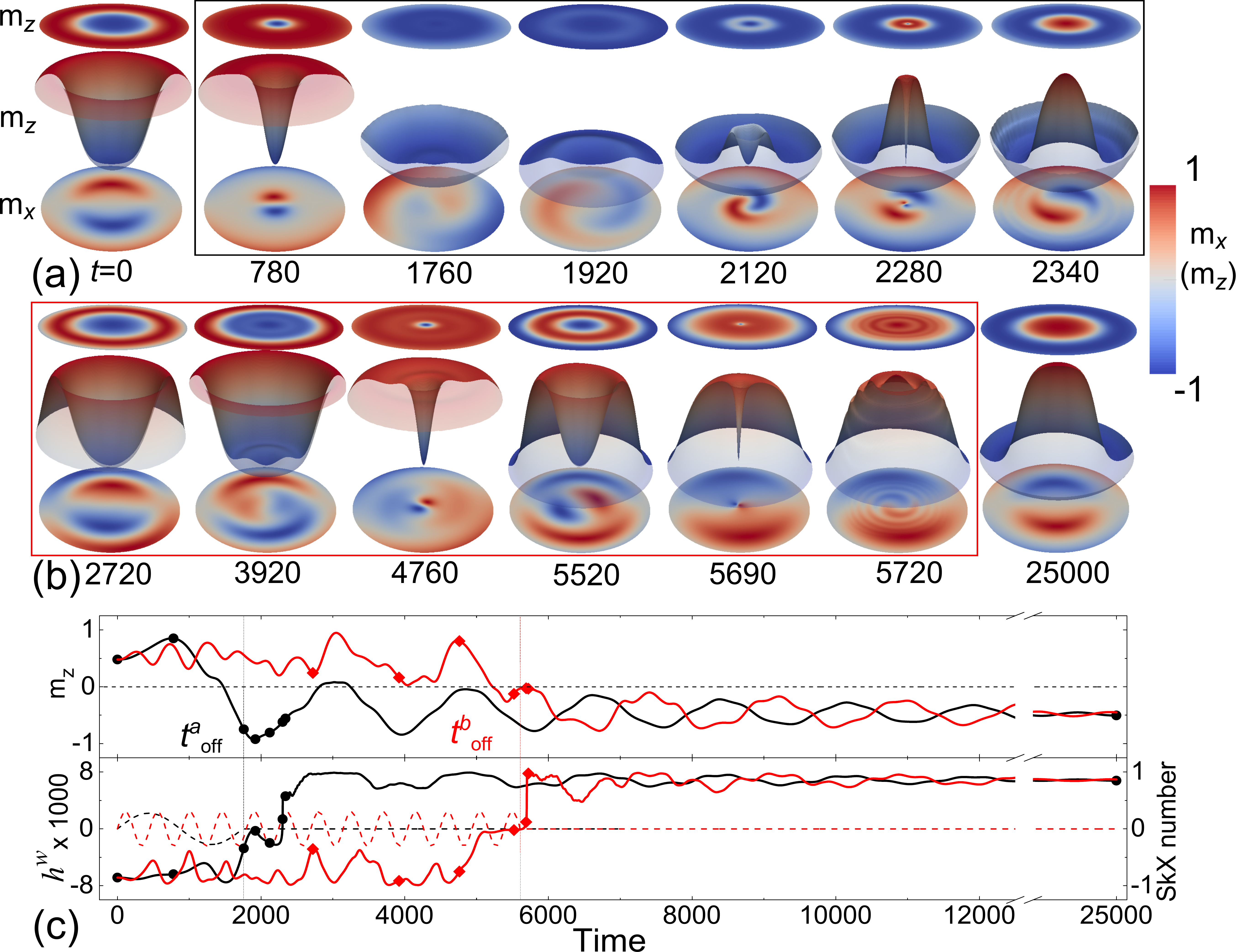}
\caption{Spin dynamics of the skyrmion excited by an AC magnetic field $h{_z^\omega} = h{^\omega} \mathrm{sin}(\omega t)$ with $h{^\omega}=2.4\times10^{-3}$ up to time 25000 ($\sim 16.5$ ns for $J=1$ meV and $S=1$).
Selected time evolution of the magnetization distribution with two examples: (a) $\omega_a=3.6 \times 10^{-3}$, $t^a_{\mathrm{off}}$ = $1750$ (multimedia view),
 (b) $\omega_b=1.4 \times 10^{-2}$, $t^b_{\mathrm{off}}$ = $5600$ (multimedia view). The color-coded $m_z$ (out-of-plane) and $m_x$ (in-plane) components of the magnetization are displayed in the first and third line, respectively, while the second line displays a perspective view (topography) with the color code of
$m_z$.  (c) Time trace of the magnetization component $m_z$ (top graph), the AC magnetic field (dashed lines), and the skyrmion number (both in bottom graph).  Black and red lines refer to the data of the two different examples, while the black circles and red prisms indicate the times corresponding to the snapshots shown in (a) and (b), respectively.
 }
\label{fig3}
\end{center}
\end{figure*}

After determining the eigenfrequencies of the skyrmion formed in a nanodot, we study the switching driven by a sinusoidal magnetic field along the $z$ direction,
i.e., $h{_z^\omega} = h^\omega \sin(\omega t)$, where $h^\omega$ and $\omega$ are the field amplitude and frequency, respectively.
Two cases of microwave-induced dynamics of a single skyrmion at $h{^\omega}=2.4\times10^{-3}$ ($ \approx 20$ mT for $J=1$ meV and $S=1$) with AC fields
of $\omega=3.6 \times 10^{-3}$ and $1.4 \times 10^{-2}$ corresponding to the breathing and mixed resonant modes, respectively, are shown in Fig.~\ref{fig3}.
Figures~\ref{fig3}(a) and (b) present serial snapshots, while the corresponding $m_z(t)$ and $h^\omega(t)$ are shown in Fig.~\ref{fig3}(c) by black and red curves for the two frequencies.
The initial skyrmion state ($t=0$) is obtained by relaxing the system with maximum $|\mathrm{d}m/\mathrm{\mathrm{d}}t|<10^{-6}$.
In the initial state, the skyrmion number $-0.86$ and the $z$ component of the magnetization $m_z$ of the skyrmion core are negative, while $m_z$ at the edge of the nanodot is positive.
In the first scenario, shown in Fig.~\ref{fig3}(a) [See also multimedia view of Fig. 3(a)], the breathing mode is excited by applying an AC field with $\omega_a=3.6 \times 10^{-3}$,
which is switched off at $t^a_{\mathrm{off}}$ = $1750$, as indicated by the black dotted line in Fig.~\ref{fig3}(c).  As a response to the AC magnetic field, the skyrmion starts to breathe,
i.e., the skyrmion core expands when the field points to the negative $z$-direction and shrinks if the direction of applied magnetic field is positive. The core thus becomes smaller
than in the initial state ($t=0$) after the first positive half period ($t=780$).  After one full period ($t=1760$), the skyrmion core expands over the entire dot, the edges become negative, and
a transient uniform state is reached. Afterwards, the tilted spins at the edge of the disk circulate to the inside of the dot
and propagate to the center as shown in Fig.~\ref{fig3}(a) at times 1920--2120.
At the same time, $m_z$ of the core begins to increase and an upward core is nucleated at $t=2340$
[Fig.~\ref{fig3}(a)].
Subsequently, high-frequency radial spin waves are excited from the core center.
The switching of the skyrmion from a downward to an upward core orientation is finished. At last the system enters into a damped oscillation with the
frequency of the breathing mode up to $t \approx 25000$, as shown by the black line in Fig.~\ref{fig3}(c).

Now we turn to the simulation results of the mixed mode case, $\omega_b=1.4 \times 10^{-2}$, with $t^b_\mathrm{off} = 5600$,
shown in Fig.~\ref{fig3}(b) [See also multimedia view of Fig. 3(b)] and by the red curves in (c). By applying the AC magnetic field, both the spins at the edge of the structure and around the skyrmion core are excited in the mixed mode, as shown at $t=2720$. After $t=3000$, the lower-frequency breathing mode starts and
then dominates the dynamic properties. The most expanded core is obtained at $t=3920$ with a ring of tilted spins around the middle,
similar to the case of $t=1920$ and $t=2120$ in Fig.~\ref{fig3}(a), however, no reversed core is formed for these field parameters.
After the core has shrunk to a smaller size at $t=4760$, it does not recover to the previous size because of a compression resulting from
the excitation of the edge area ($t=5520$).  Instead, the core shrinks even further, and immediately thereafter, the down-core vanishes and
is replaced by an up-core, while spin waves are emitted from the core area (see $t=5720$). After switching off the AC field at $t=5600$,
the reversed skyrmion starts to dissipate the energy via both the breathing and mixed modes.

Similar to the vortex core reversal with out-of-plane fields,~\cite{Yoo2012, Wang2012, Ruckriem2014, Noske2015} the reversal processes for
the two frequencies are symmetric, and the switching mechanisms are similar to the radial-spin-wave-mode-assisted switching for a vortex core.~\cite{Yoo2012}  At the lower frequency,
which corresponds to the breathing mode, the skyrmion first switches to a transient uniform state via axial core expansion to the edge,
then a new core is formed from the center. For the mixed mode excitation at the higher frequency, the reversed skyrmion results from the shrinking of the core
with the help of spin excitations from the edge region.  High-frequency spin waves are emitted during the core reversal,
however, a spin-wave assisted re-reversal, i.e., a switching back to the original state by a spin wave,
as it has been observed in vortex core reversal,~\cite{Yoo2012, Wang2012, Ruckriem2014} has not been found here.
Our simulation results also show that the switching time and switching field for a skyrmion-core reversal are similar to those needed for a vortex-core reversal.

\begin{figure}[tbhp]
\begin{center}
\includegraphics[scale=0.35]{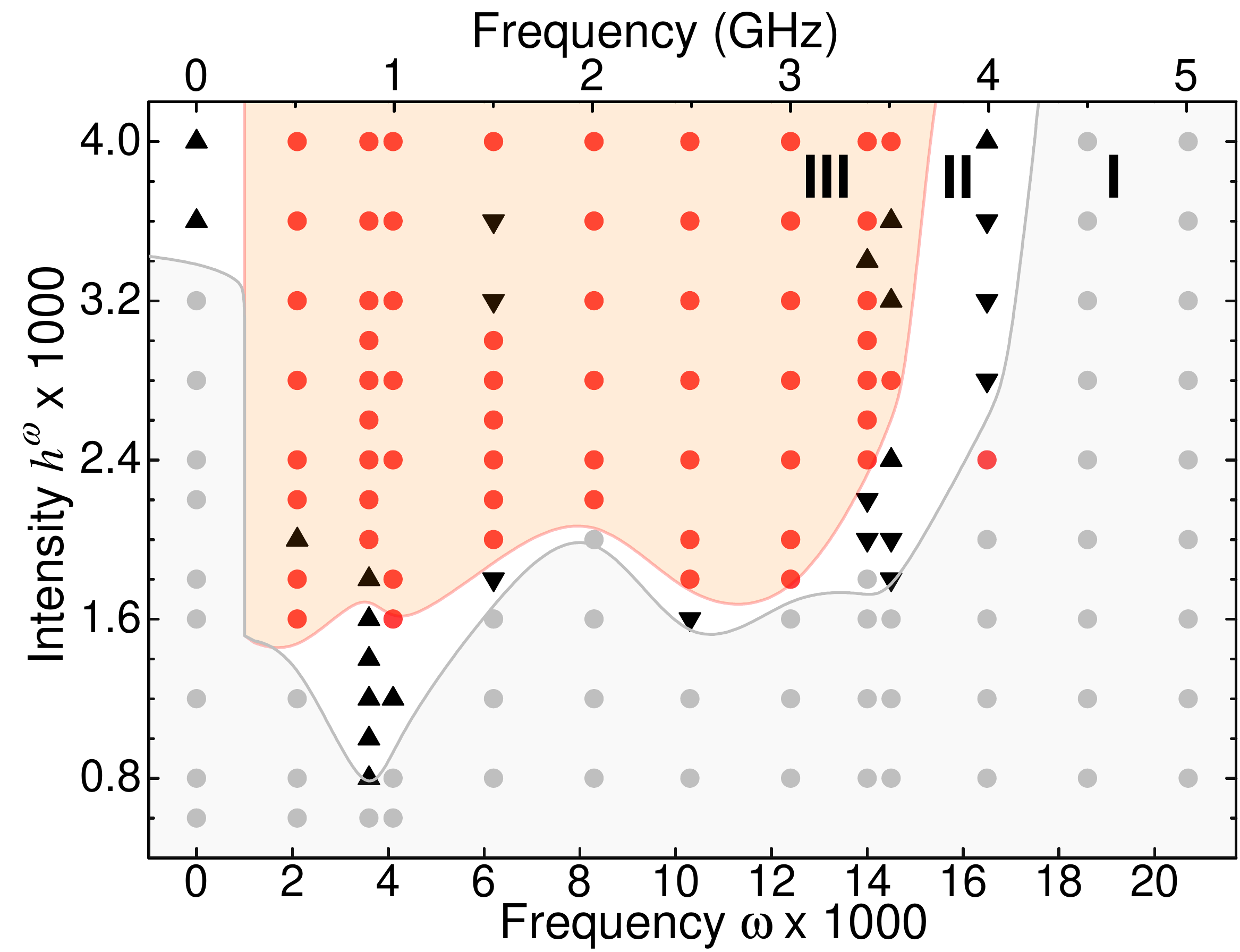}
\caption{Phase diagram of the skyrmion switching induced by a field $h{_z^\omega} = h{^\omega} \mathrm{sin}(\omega t)$ as a function of $h{^\omega}$ and $\omega$.
Red dots indicate skyrmion reversal, gray dots mean no switching,
up-pointing and down-pointing triangles skyrmion switching to uniform and distorted states, respectively.
 }
\label{fig4}
\end{center}
\end{figure}

Besides the duration time, the skyrmion reversal process also depends on the frequency $\omega$ and the AC field amplitude $h^\omega$.  Figure \ref{fig4} shows the switching phase diagram as a function of $\omega$ and $h^\omega$
with a maximum temporal duration $t=10000$ ($\sim 6.6$ ns for $J=1$ meV and $S=1$).
The diagram can be divided into three different regions. The first is the weak-susceptibility region I (Gray symbols), in which a skyrmion core never switches for AC field duration up to $t=10000$,
mainly because the field is not strong enough with respect to the dynamic susceptibility of the skyrmion.
In the second region (II) a skyrmion converts to either the uniform state (up-pointing triangles) or the distorted state (down-pointing triangles). The distorted state is formed directly from the skyrmion due to
the instability of the core under the AC field with intermediate intensities, similar to the case of magnetic vortices.~\cite{Pylypovskyi2013}  The boundary between regions I and II reflects the susceptibility spectrum of the skyrmion, cf.\ Fig.\ \ref{fig1}, in such a way that at around the eigenfrequencies of the system switching to the uniform or distorted states occurs at lower field amplitudes than at other frequencies.
In the third region (III, red), a skyrmion core can be reversed by the AC field, as had been shown for two instances with $t^a_\mathrm{off} = 1750$ and $t^b_\mathrm{off}=5600$ in Fig.~\ref{fig3}(a) and Fig.~\ref{fig3}(b), respectively.
Comparing to the results for a static magnetic field ($\omega=0$), where in this amplitude regime the final state is either unchanged or uniform depending on whether $h^\omega$ is below or above about $3.6\times10^{-3}$, it is obvious that an AC magnetic field within a certain frequency range substantially helps to reverse the skyrmion core.

In summary, we have studied the nonlinear skyrmion dynamics in a circular magnetic nanodot induced by a vertical oscillating magnetic field using micromagnetic simulations.
We found that in a certain frequency window a fast skyrmion switching can be achieved for field amplitudes that do not lead to a reversed skyrmion under static conditions.
We presented two examples of skyrmion switching in detail, in which the excitation frequencies corresponded to the breathing and mixed modes of the system.
Under excitation with the breathing-mode frequency, the skyrmion reverses via a transient uniform state, while exciting it with the mixed-mode frequency spin waves from the edge area assist the switching.
Our results show that a skyrmion core can be reversed within nanoseconds by means of microwaves perpendicular to the skyrmion plane.

B.Z. gratefully acknowledges funding by the China Scholarship Council.
W.W. and M.B. acknowledge financial support from EPSRC's DTC grant EP/G03690X/1.
The High Performance Computing Center at the Freie Universit\"{a}t Berlin (ZEDAT) is acknowledged for computational time.

\bibliographystyle{apsrev4-1}
%


\end{document}